\documentclass[conference]{IEEEtran}
\IEEEoverridecommandlockouts
\usepackage{cite}
\usepackage{amsmath,amssymb,amsfonts}
\usepackage{algorithmic}
\usepackage{graphicx}
\usepackage{textcomp}
\usepackage{circuitikz}
\usepackage{xcolor}
\def\BibTeX{{\rm B\kern-.05em{\sc i\kern-.025em b}\kern-.08em
    T\kern-.1667em\lower.7ex\hbox{E}\kern-.125emX}}
\begin{document}

\title{Hardware Trojans in Power Conversion Circuits}

\author{\IEEEauthorblockN{1\textsuperscript{st} Jacob Sillman}
\IEEEauthorblockA{\textit{ECE Department} \\
\textit{UC Davis}\\
Davis, CA \\
jlsillman@ucdavis.edu}
\and
\IEEEauthorblockN{2\textsuperscript{nd} Ajay Suresh}
\IEEEauthorblockA{\textit{ECE Department} \\
\textit{UC Davis}\\
Davis, CA \\
ajsuresh@ucdavis.edu}
}

\maketitle

\begin{abstract}

This report investigates the potential impact of a Trojan attack on power conversion circuits, specifically a switching signal attack designed to trigger a locking of the pulse width modulation (PWM) signal that goes to a power field-effect transistor (FET). The first simulation shows that this type of attack can cause severe overvoltage, potentially leading to functional failure. The report proposes a solution using a large bypass capacitor to force signal parity, effectively negating the Trojan circuit. The simulation results demonstrate that the proposed solution can effectively thwart the Trojan attack. However, several caveats must be considered, such as the size of the capacitor, possible current leakage, and the possibility that the solution can be circumvented by an adversary with knowledge of the protection strategy. Overall, the findings suggest that proper protection mechanisms, such as the proposed signal-parity solution, must be considered when designing power conversion circuits to mitigate the risk of Trojan attacks.

\end{abstract}

\section{Introduction}
Hardware Trojans are a serious threat to the security and reliability of electronic systems, and power conversion circuits are particularly vulnerable due to their closed-loop, analog nature. Among hardware Trojans, power block hardware Trojans are especially dangerous since they are designed to be small in size, have low-power operation, and can be triggered with a small number of gates. This makes them difficult to detect and almost impossible to remove without damaging the circuit.

This report aims to present a study of the effects of analog Hardware Trojans on power conversion circuits, with a specific focus on power block Hardware Trojans. As noted, these Trojans are designed to be small in size, low power, and hard to detect, making them a severe threat to power conversion circuits. Power block hardware Trojans are typically placed inside the power generation block, where they can live with minimal leakage of side-channel information due to the large electromagnetic fields and hot operation. Furthermore, since power generation circuits are closed-loop and analog, they are challenging to test with conventional methods, and the only well-monitored ports of a power block are the input and output power.

For the overall impact and motivation behind such a Trojan, this report considers the threat model of untrusted third-party intellectual property (3PIP), untrusted system-on-chip (SoC) developers, and untrusted foundries. Power intellectual property is easy to reverse engineer on the side of the foundry, making it challenging to obfuscate and protect against hardware Trojans.

\section{Threat Models}

\subsection{Untrusted Third Party IP}

The first threat model pertains to an untrusted third-party intellectual property (3PIP) that designers might have acquired, with a built-in trojan. In this case, designers use a third-party IP to reduce the design complexity and development cost. However, they have no control over the design of the IP, and if the third-party intentionally inserts a Trojan into the IP, it will remain unnoticed by the designers. This poses a severe threat to the entire system and can result in significant damages. 

\subsection{Untrusted System on Chip Developer}
The second threat model deals with an attacker who maliciously adds a Trojan to the system during the design phase. This attacker exists within the design house and has the necessary knowledge and expertise to insert a Trojan while designing the circuit. Since the attacker has access to the entire design, they can inject the Trojan in any part of the circuitry, making it more challenging to detect. 

\subsection{Untrusted Foundry}
The third threat model involves an untrusted foundry that deviates from the netlist and adds a Trojan during the fabrication process. In this case, the attacker has access to the hardware, and they can modify the circuitry to insert a Trojan. Since the Trojan is inserted during the fabrication process, it becomes more challenging to detect as it is masked by the hardware's natural variations. 

\section{Experimental Setup}

\subsection{Trojan}

For the scope of this paper, we only modelled the trigger circuit of the trojan. The circuit that generates the trigger signal can take a variety of different forms. Some condition circuits are digital in nature, and can take the form of sequential or combinational logic \cite{3}. These circuits still need to be small in area and gate-number in order to escape detection. There are other types of condition circuits that are analog in nature, such as large-delay trojans that combine the effects of gate-oxide leakage current and Miller capacitance to create delay signals up to 2 days in length\cite{2}. 

The trojan trigger circuit being tested is a switching signal attack that is designed to trigger a locking of the pulse width modulation (PWM) signal that goes to a power field-effect transistor (FET). The purpose of this attack is to cause a system to either overvolt or disable entirely. This attack is accomplished by using an OR/NOR gate with both the PWM signal and the trojan signal as inputs, with the output going to the gate of the power FET.

When the trojan is triggered, the power FET is locked with either a high or low voltage, preventing the proper voltage regulation and resulting in potential system damage. This type of attack can be particularly damaging in the context of an IC buck converter, where the voltage is regulated and converted for a variety of applications. The use of a power FET in the attack also indicates that the trojan is designed to target power conversion circuits specifically, further underscoring the potential impact of such an attack.

\begin{figure}[htbp]
\centering
\includegraphics[width=\columnwidth]{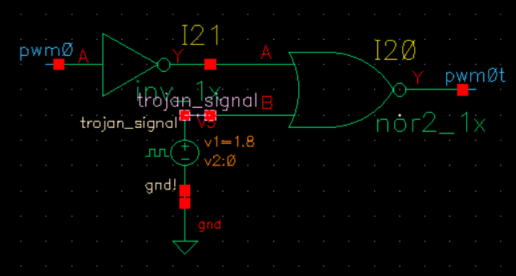}
\caption{Logic Trigger for Trojan}
\label{fig}
\end{figure}

The simulated trigger circuit contains a total of 2 gates, 7 transistors (could be implemented with XOR as well). This makes it particularly difficult to detect through conventional testing methods, such as inserting test patterns to assess loop functionality. Additionally, power generation circuits are typically closed-loop and analog, which limits the ability to insert test patterns to evaluate system functionality.

Overall, a well-designed power-management Hardware Trojan can have a small footprint, a rare trigger condition, and be difficult to detect without destructive testing. This highlights the potential threat of such attacks and underscores the need for continued research and development of methods to detect and prevent Hardware Trojans in power conversion circuits.
\subsection{DC-DC Circuit}

To test the effects of the Trojan, a DC-DC converter topology is simulated (as shown above). It has a load current of 10 mA and operates at 93.3\% efficiency. The output ripple of the converter is 23.8 mVpp. The converter is designed using an inductor with a value of 55.5 uH and an ESR of 0.777 ohms, along with a capacitor with a value of 40 nF and an ESR of 0.358 ohms. The power FETs used in this circuit have a fanout ratio of 4.

\begin{center}
     \begin{circuitikz}[american, scale = 0.7, transform shape]
         \ctikzset{tripoles/mos style/arrows}
         \ctikzset{transistors/arrow pos=end}
         \draw (0,-1) node (m1) [nmos] {N$_1$};
         \draw (0,2.5) node (m2) [pmos] {P$_1$};
         \draw (m2.D) -- (m1.D) {};
         \draw (0,0.75) to[L, *-, l=L] ++ (2,0) to[R, l={R$_{ESRL}$}] ++(2,0) to[C, l=C, *-] ++(0,-1.5) to[R, l={R$_{ESRC}$}] ++(0,-1.5){};
         \draw (4,0.75) to[short, -*] ++(2,0) to[R, l={R$_L$}] ++(0,-3){};
         \draw (6,0.75) -- ++(2,0) to[isource, l={I$_L$}] ++(0,-3){};
         \draw (m1.S) |- (4,-2.25) to[short,*-*] ++(2,0) -- ++(2,0) node[ground]{};
         \draw (m2.S) node[tground]{};
         \draw (m2.S) node[above]{V$_{supply}$};
         \draw (-3,-1) node (in1) [ieeestd not port, scale = 0.5]{};
         \draw (-1.75,-1) node[ieeestd not port, scale = 1]{};
         \draw (-3,2.5) node (in2)[ieeestd not port, scale = 0.5]{};
         \draw (-1.75,2.5) node[ieeestd not port, scale = 1]{};
         \draw (in2.in) node[left]{PWM$_0$};
         \draw (in1.in) node[left]{PWM$_1$};    
     \end{circuitikz}
     
     Buck Converter Circuit Diagram
 \end{center}

\section{Key Findings}

During the simulation, the Trojan was inserted into the PMOS (upper transistor) PWM signal, resulting in a lock-down of the PMOS signal to a low voltage state. As a result, the PMOS remained on at all times, causing overvoltage in the system. The simulation showed that at a load current of 10mA, the system voltage reached as high as 1.2V. This overvoltage could potentially push a processor to a voltage corner, leading to functional failure. The waveform obtained from the simulation clearly depicts the overvoltage caused by the Trojan attack (Figure 2). Table 1 shows an extrapolation of the simulation results to show the various other effects that a similar attack could have on this circuit.

\begin{figure}[htbp]
\centering
\includegraphics[width=\columnwidth]{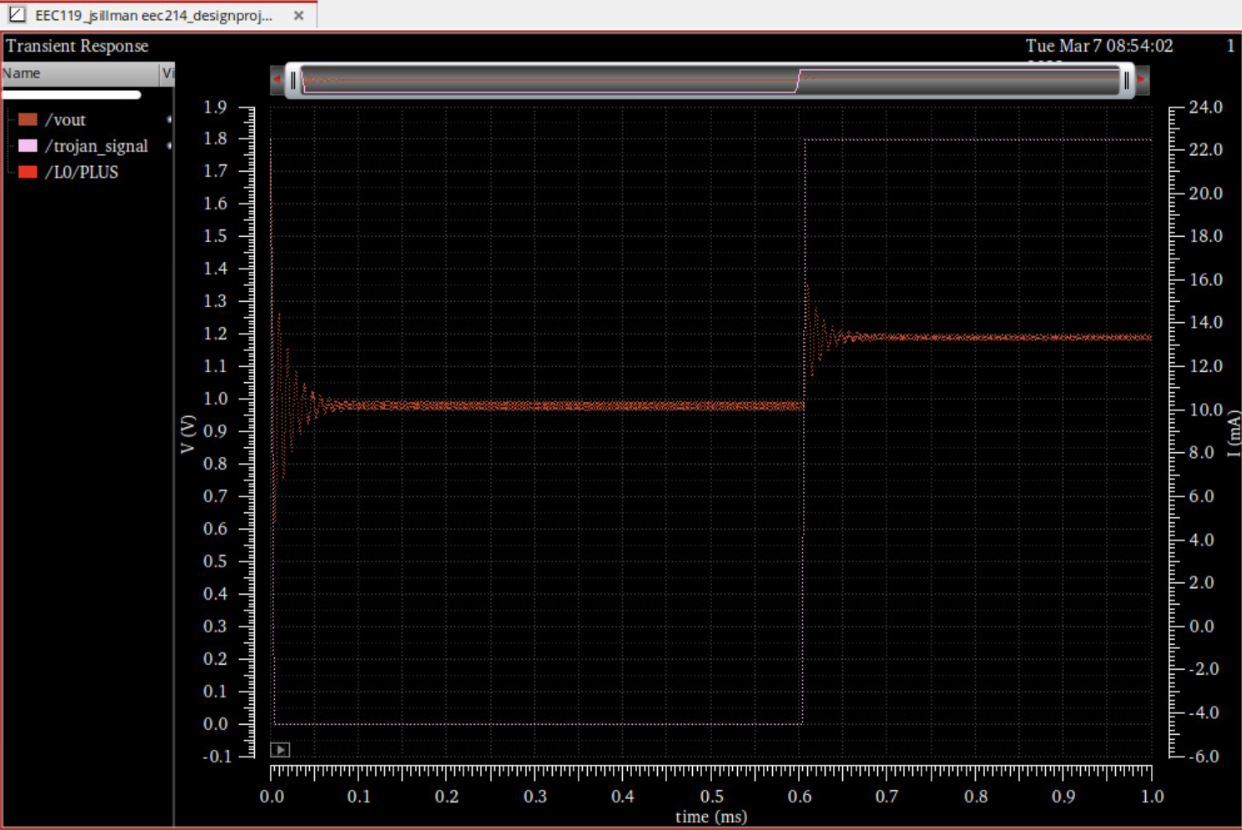}
\caption{Shift of operating voltage from 1V to 1.2V}
\label{fig}
\end{figure}

\begin{table}[htbp]
\caption{}
\begin{center}
\begin{tabular}{|c|c|c|c|}
\hline
\textbf{}&\multicolumn{3}{|c|}{\textbf{Results}} \\
\cline{2-4} 
\textbf{S.No.} & \textbf{\textit{PMOS Gate}}& \textbf{\textit{NMOS Gate}}& \textbf{\textit{Result}} \\
\hline
1& Not Locked & Locked Low & NMOS open circuit, severely overvolts\\
\hline
\hline
2& Not Locked & Locked High & NMOS short circuit, disables power block\\
\hline
\hline
3& Locked Low & Not Locked & PMOS short circuit, overvolts system\\
\hline
\hline
4& Locked High & Not Locked & PMOS open circuit, disables power block\\
\hline

\end{tabular}
\end{center}
\end{table}

\section{Impacts}

As shown clearly through the above simulation, hardware trojans can cause subtle changes in the supply voltage of electronic systems that can have significant impacts on a variety of applications that use power converter topologies in their systems similar to the one simulated above. The increase in supply voltage, prolonged over a period of time can lead to Negative Bias Temperature Instability (NBTI) phenomena, which can degrade the performance of electronic components over time\cite{1}. NBTI can lead to a decrease in transistor speed and an increase in leakage current, which can cause a degradation in circuit performance and reliability.\cite{1}

In sensors that are very sensitive to subtle changes in supply voltage, such as those used in critical applications like medical devices, Hardware Trojans can have severe consequences. For example, pacemakers rely on highly precise sensors that can detect subtle changes in voltage levels to regulate the heart's rhythm. If a hardware Trojans were to compromise the pacemaker's sensors, it could cause the device to malfunction or provide inaccurate readings, leading to severe health consequences such as heart attacks or even death.

Furthermore, sensors used in medical devices often operate in harsh environments that can cause significant fluctuations in the supply voltage. These fluctuations can exacerbate the impact of Hardware Trojans on the reliability of the device. 

\section{Potential Mitigation}

A second simulation was conducted to implement and test a potential mitigation to this kind of Hardware Trojan. The proposed solution involved using a large capacitor to force signal parity. The capacitor was placed at the root of PWM signal generation, which could be at the output of a VCO or comparator network. The other end of the capacitor was tied to the same net as the power FET gate. This ensured that the parity of the true PWM signal generation was checked and what voltage was being measured at the gate of the FET (Figure 3).

\begin{figure}[htbp]
\centering
\includegraphics[width=\columnwidth]{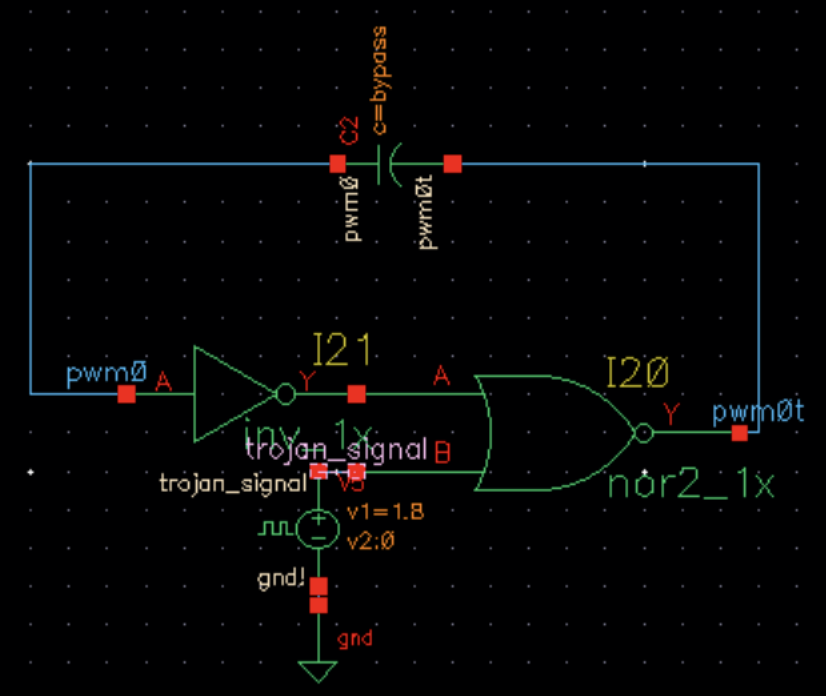}
\caption{Trigger circuit thwarted with parity capacitor @ 500 pF}
\label{fig}
\end{figure}

The simulation showed that if the voltage at the gate of the FET and the PWM were the same, no current would flow through the capacitor, and normal functionality would be maintained. However, if the PWM and the gate voltage were opposite in phase for any amount of time, current would flow through the capacitor, alter the gate voltage, and trigger the Trojan. This effectively negated the Trojan circuit, turning the inserted OR/NOR gate into a phase-shift network or delay network (Figure 4).

\begin{figure}[htbp]
\centering
\includegraphics[width=\columnwidth]{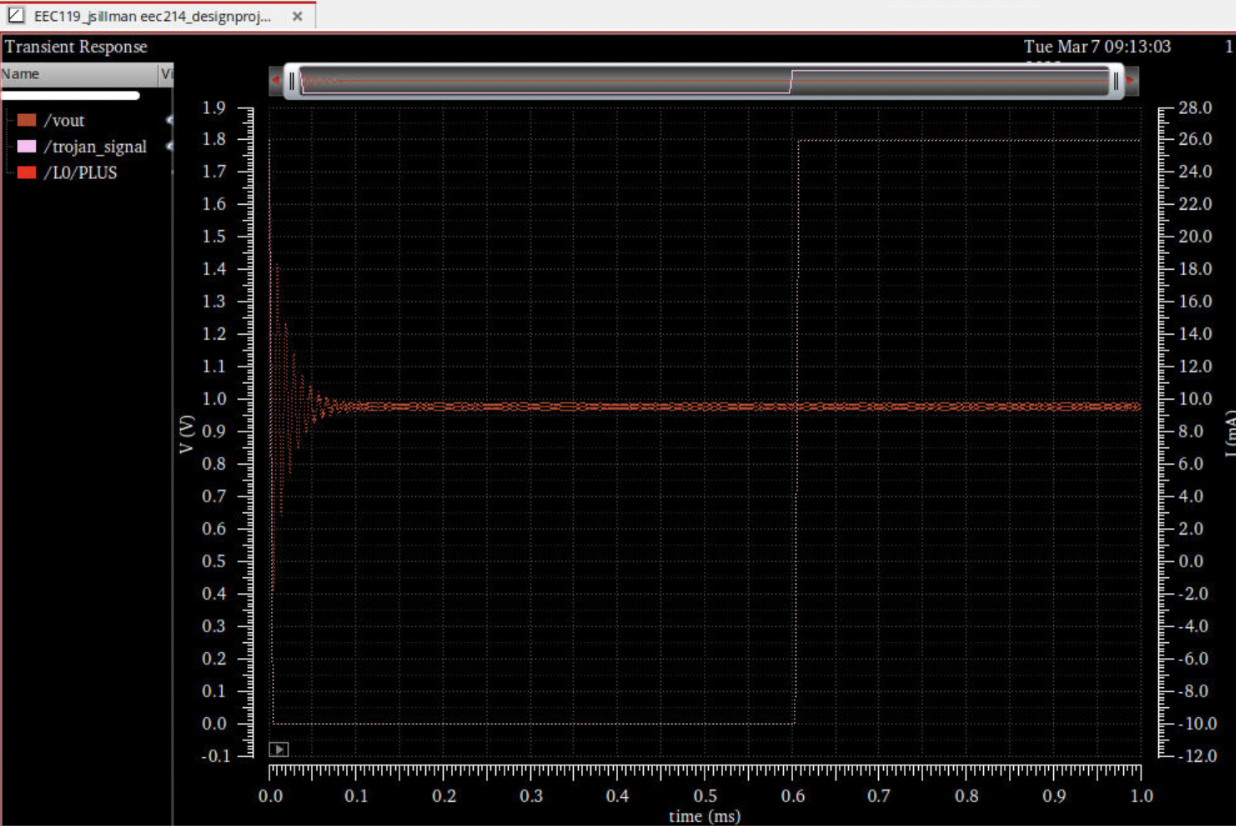}
\caption{Effect of parity 500 pF capacitor}
\label{fig}
\end{figure}

It should also be noted that this kind of signal-parity solution could be applied to any number of switching topologies. However, there are some caveats to consider. Firstly, the capacitor must be large enough to pass the PWM signal. For the simulation, 500 pF was required at 1 MHz switching frequency, which was not possible to fabricate, so an external capacitor was suggested. Secondly, the travel time of the PWM signal may automatically incur some phase shift between the plates of the capacitor. This could impact efficiency metrics due to a slight change in duty cycle measured at the FET gate, even without Trojan interference. This could be preemptively considered during design. Finally, an adversary with knowledge about this kind of protection strategy could fairly easily circumvent it by placing their gate on either side of the capacitor.

\section{Conclusions and Evaluation}

In conclusion, this report has demonstrated the potential impacts of a Trojan attack on power management circuits. The simulation results clearly show how a simple OR/NOR gate can be used to trigger a locking of the PWM signal and cause overvoltage or system failure. The results also indicate that such attacks can be difficult to detect through conventional testing methods, which highlights the need for more sophisticated protection strategies.

The proposed solution of using a large bypass capacitor to force signal parity appears to be an effective countermeasure against this type of Trojan attack. However, this approach comes with its own set of limitations and caveats, such as the need for a large enough capacitor to pass the PWM signal and the possibility of current leakage impacting efficiency metrics.

Overall, this report highlights the importance of protecting power management circuits against Trojan attacks and the need for more advanced protection strategies. As technology continues to advance, it is crucial that researchers and designers remain vigilant in identifying potential vulnerabilities and developing effective countermeasures to ensure the reliability and security of critical systems.

\section{Project Contributions}

\begin{figure}[htbp]
\centering
\includegraphics[width=\columnwidth]{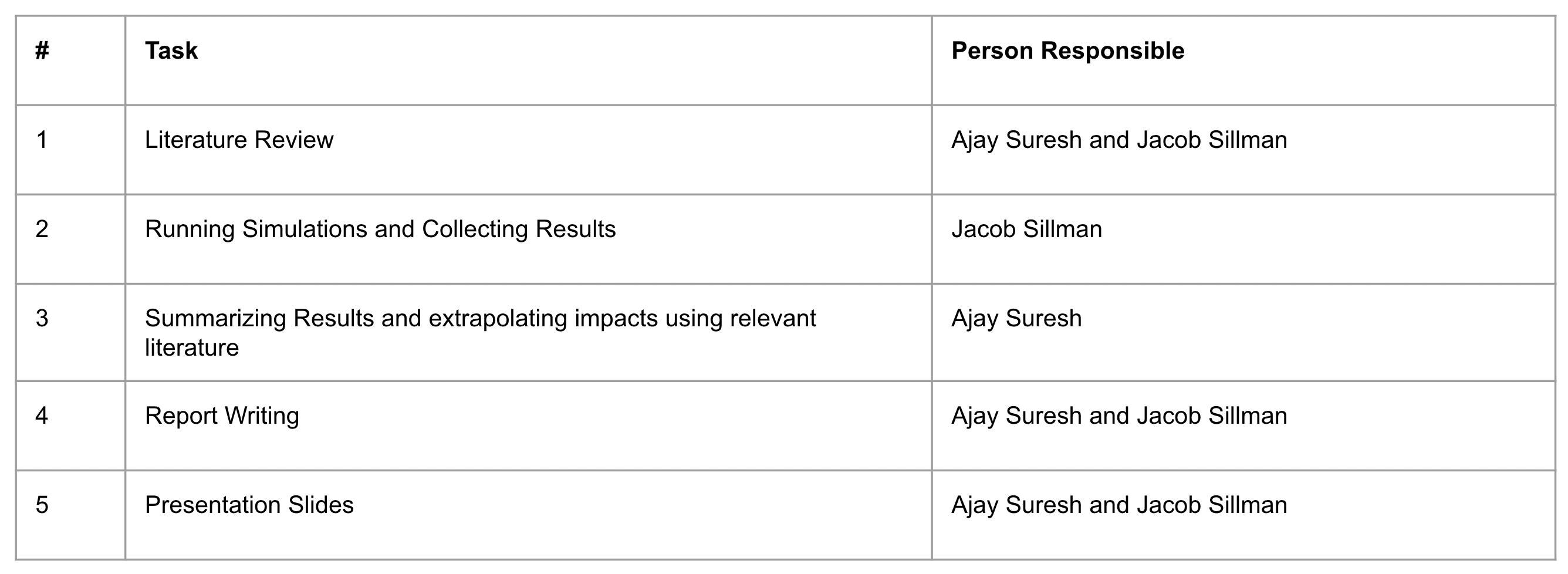}
\caption{Distribution of Work}
\label{fig}
\end{figure}

\end{document}